\documentclass[conference,conference,compsocconf]{IEEEtran}

\usepackage[cmex10]{amsmath}
\usepackage{cite}
\usepackage{epsfig}
\usepackage{color}
\usepackage[usenames,dvipsnames]{xcolor}
\usepackage{nopageno} 
\usepackage{fancyhdr}
\usepackage[absolute]{textpos} 
\usepackage{float}
\usepackage{graphicx} 
\usepackage[T1]{fontenc}
\usepackage{textcomp}
\usepackage{caption}
\usepackage{subfigure}
\usepackage{array}

\begin{document}

\bstctlcite{IEEEexample:BSTcontrol}

\title{Enhanced HARQ for Delay Tolerant Services in Mobile Satellite Communications}
\author{\IEEEauthorblockN{Rami Ali Ahmad, J{\'e}r{\^o}me Lacan}
\IEEEauthorblockA{University of Toulouse, ISAE/DMIA \& T{\'e}SA\\
Toulouse, France\\
Email: rami.aliahmad@isae.fr,\\ jerome.lacan@isae.fr}
\and
\IEEEauthorblockN{Fabrice Arnal, Mathieu Gineste}
\IEEEauthorblockA{Thales Alenia Space\\
Toulouse, France\\
Email: fabrice.arnal@thalesaleniaspace.com,\\ mathieu.gineste@thalesaleniaspace.com}
\and
\IEEEauthorblockN{Laurence Clarac}
\IEEEauthorblockA{CNES\\
Toulouse, France\\
Email: laurence.clarac@cnes.fr}}

\maketitle
\begin{abstract}
The objective of our paper is to improve efficiency (in terms of throughput or system capacity) for mobile satellite communications. In this context, we propose an enhanced Hybrid Automatic Repeat reQuest (HARQ) for delay tolerant services.  
Our proposal uses the estimation of the mutual information. We evaluate the performance of the proposed method for a land mobile satellite channel by means of simulations. Results are compared with those obtained with a classical incremental redundancy (IR) HARQ scheme. The technique we propose, shows a better performance in terms of efficiency while maintaining an acceptable delay for services.

\end{abstract}

\IEEEpeerreviewmaketitle


\section{Introduction}\label{sec:Intro}

For many years, mobile satellite communications services are challenging and very costly. Land mobile satellite (LMS) channels are highly affected by important propagation impairments (time-selective channel) and inter/intra system interference, that both cause unstable and low signal to noise ratio (SNR), which the receivers have to cope with. This usually implies strong limitations on the delivered service throughput, whatever the considered satellite frequency band.

Our objective is to propose a mechanism, which improves the efficiency of link usage while providing an appropriate service to applications. The targeted services (data transfer from sensors, messages for aeronautical services, etc.) are assumed to be tolerant to delay. For example some aeronautical services define delay requirement for the delivery of 95\% of messages \cite{air_traffic}.

To deal with problems caused by link characteristics in mobile satellite communications, there are many solutions. One of these solutions is to use pure Forward Error Correction (FEC), which can makes the message very robust. However using FEC is not sufficient sometimes, due to the highly varying channel, where its quality changes dramatically. This make it difficult sometimes to decode the message, even if the used code is very robust. Alternatively, Automatic Repeat reQuest (ARQ) can be used as a solution to deal with channel variations, where transmitter attempts many retransmissions in case of unsuccessful decoding. ARQ messages without FEC, are not so robust to ensure a reliable communication. 

Hybrid ARQ (HARQ) protocols are used in most of recent terrestrial wireless communication systems. Worldwide interoperability for microwave access (WIMAX) and long term evolution (LTE) are examples of these systems that use HARQ \cite{overview_arq}. There are many types of HARQ techniques. HARQ type I is called chase combining (CC). The receiver asks the transmitter for the transmission of the same packet of coded data. At the receiver, the decoder combines the multiple copies of the transmitted packet weighted by the SNR received. HARQ type II, is also called incremental redundancy (IR). Unlike the previous method, which transmits simple repetitions of the same encoded packets, IR technique transmits additional redundant information in an incremental way if the decoding does not succeed from the first transmission. If each retransmission packet is self-decodable this scheme is called type III HARQ \cite{panasonic}. Classical IR HARQ, which transmits a fixed number of bits at each transmission, is not optimal from the efficiency point of view even if it can improve the decoding delay especially for high values of SNR. However, choosing optimal values for number of bits to be sent at each transmission can improve the system performance and throughput level. Many papers has proposed methods to choose an optimal number of bits at each transmission, a theoretical basis for a parity bits selection by means of a risk-sensitive optimal control is established in \cite{Denic}. The idea of random transmission assignments of the mother code bits was introduced in \cite{Soljanin}. An optimized IR HARQ schemes based on punctured LDPC codes over the BEC was proposed in \cite{Andriyanova}. 

Contrarily to classical IR HARQ scheme, the main idea of our enhanced HARQ technique is to estimate the average number of bits to be transmitted at each transmission to decode the codeword with a targeted probability. This probability depends on the application/service (delay constraints). This technique uses the mutual information to predict the mean number of bits needed for each transmission. It uses the knowledge of the statistical distribution of the channel attenuation.  
Our enhanced HARQ transmission proposal is simulated in a satellite communications environment, where an LMS channel and a long Round-Trip time are considered. 

The remainder of this paper is organised as follows. Section II describes the channel capacity and the LMS channel model. We present classical IR HARQ in Section III. In Section IV, we propose an enhanced HARQ model for delay tolerant services in mobile communications. We present the results of simulations and we compare the performance of both techniques (classical IR and enhanced HARQ) in Section V. 
We conclude our study in Section VI.

\section{Channel modelling}\label{sec:cahn_mod}
In our study on mobile satellite communications, we considered an LMS channel to model this environment \cite{LMSPropa}\cite{LMSStatmodel}. In the following, we explain how to compute the capacity of this channel. 

\subsection{Channel capacity and Mutual Information}\label{sec:chan_capa}

Channel capacity $C$ quantifies the maximum achievable transmission rate of a system communicating over a band-limited channel, while maintaining an arbitrarily low error probability.
It corresponds to the maximum of the mutual information between the input and output of the channel, where the maximization is done with respect to the input distribution. Mutual information (MI) measures the information that input of the channel ($X$) and output of the channel ($Y$) share. It is a key parameter in our approach, as it will be used to calculate the mean number of bits needed to decode a message at each transmission with a given probability. 

Given the channel input symbol $x_i$, its energy ${E_s}_i$, a realisation of noise $n_i$ (which has a Gaussian distribution with variance $\frac{N_0}{2}$) and the channel attenuation $\rho_i$, the channel output symbol $y_i$ can be written as:
\begin{equation}\label{input_output}
\begin{aligned}
y_i=\rho_i \sqrt{{E_s}_i}x_i+n_i.
\end{aligned}
\end{equation}

For an equally distributed input probability, the MI corresponds to the capacity of the channel, which is AWGN, can be calculated by the following equation \cite{turbo}:


\begin{equation}\label{capacity_LMS}
\begin{aligned}
MI\left(\frac{E_s}{N_0}\right) &= \log_2(M)-\frac{1}{M (\sqrt{\pi})^N} \times \\ &\sum_{m=1}^M \underbrace{\int_{-\infty}^{+\infty}\dotsi \int_{-\infty}^{+\infty} }_\text{N times} \exp(-|t|^2)\\ &\times \log_2\left[\sum_{i=1}^M\exp(-2t.d_{mi}-|d_{mi}|^2)\right]\,dt, \,\,
\end{aligned}
\end{equation}

where:
\begin{itemize}
\item $M$ is the modulation order;\\
\item $N$ is the space dimension that depends on the used modulation ($N=2$ for any PSK-based modulation with more than 2 states);\\
\item $t$ is the integration variable of dimension $N$ ($t$ = ($t[1],...,t[N]$));\\
\item $d_{mi} = \sqrt{\frac{E_s}{N_0}}(x_m-x_i)$ ($x_i$ it is an input symbol).\\
\end{itemize}

For the rest of the paper, we define $MI_{req}$ as the average MI per bit required to decode a codeword at a given probability expressed in Word Error Rate (WER). $MI_{req}$ can be calculated using the mutual information function \eqref{capacity_LMS} and the performance curves giving the WER versus $E_s/N_0$ of the used modulation/coding scheme \cite{Dolinar}. By using these performance curves, we can deduce the $E_s/N_0$ necessary to obtain a given WER. Then we use this value of $E_s/N_0$ in the mutual information function \eqref{capacity_LMS} to obtain the mutual information required to decode a codeword at this given WER. The prediction of performance of the WER based on MI is quite classical, and has been described and validated in \cite{predic_perf}. 


From now on, to compute the mutual information we use \eqref{capacity_LMS}, $\frac{E_s}{N_0}$ is replaced by $\rho_i ^2 \frac{E_s}{N_0}$, where $\frac{E_s}{N_0}$ is the average energy per symbol.

%

\subsection{LMS Channel}\label{sec:LMS_Chan}

One of the reference propagation models for LMS channel is a statistical model based on a three state Markov chain \cite{LMSPropa}.
This model considers that the received signal originates from the sum of two components: the direct signal and the diffuse multipath. The direct signal is assumed to be log-normally distributed with mean $\alpha$ (decibel relative to LOS (Line Of Sight)) and standard deviation $\Psi$ (dB), while the multipath component
follows a Rayleigh distribution characterized by its average power, $MP$ (decibel relative to LOS). This model is called Loo distribution \cite{LMSStatmodel}\cite{Loo}. 
For the modelling of the LMS channel in our simulations, we use attenuation time series using a propagation simulator based on the three state channel \cite{LMSPropa}\cite{LMSStatmodel} provided by CNES. Using this tool we calculate the distribution of the probability to obtain an attenuation in the channel for a given environment. 








 \begin{figure}[t]
\centering
\captionsetup{justification=centering}
\includegraphics[scale=0.33]{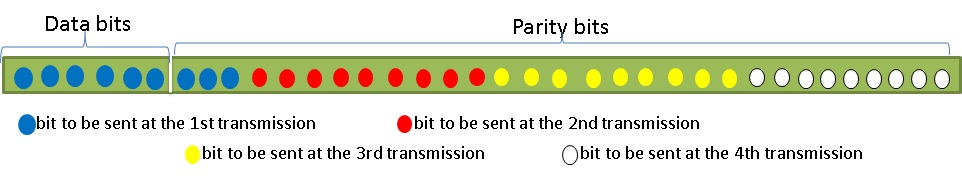} 
\caption{Example of the transmission technique considered in our simulations (Code rate 1/6, with a maximum of 4 retransmissions).}
\label{fig:puncturing}
\vspace{-15pt}
\end{figure}

\section{Incremental Redundancy HARQ}\label{sec:first_scenario}
IR HARQ can be described as follows. The sender transmits a number of bits that correspond to a given codeword. After receiving the feedback (ACK/NACK) from the receiver, the transmitter decides to no longer send bits corresponding to this codeword if an ACK is received, or to send more parity bits if a NACK is received. 
The number of bits to be sent in the next retransmission is fixed at each transmission. 
These numbers of bits are chosen without any knowledge about the channel quality or the global channel characteristics. If the cumulative received sequence can not be decoded after a maximum number of bits transmitted, the transmitter stops sending bits that correspond to this codeword. The parity bits are generated according to a coding scheme with a code rate corresponding to the maximum number of bits that can be transmitted per codeword. Bits to be sent at each transmission are part of the original codeword (mother code), leading to a different code rate at each transmission (see Figure \ref{fig:puncturing}). This technique of transmission is somehow similar to puncturing. Data bits of each codeword must be kept at the buffer of the sender as long as the codeword is still not decoded or the transmitter has not decided to end the transmission. At the receiver side, the LLRs (Log Likelihood Ratio) of the received symbols of each codeword are kept in the buffer as long as the codeword is not decoded or the transmission of the corresponding bits has not ended.
An overview of ARQ and HARQ mechanisms implemented or proposed in beyond 3rd generation systems was presented in \cite{overview_arq}.





\section{Enhanced HARQ for delay tolerant services in mobile satellite communications}\label{sec:second_scenario}
In this section, we propose an enhanced HARQ model using mutual information. 
In our scheme, the computation of the numbers of bits to be transmitted at each transmission is done in a way that insures the decoding with a probability. These probabilities are predefined for each transmission and chosen according to delay constraints of the application.


In the following, we explain how to proceed to compute the number of bits required to decode a codeword with a predefined probability, supposing the knowledge of the global statistics of the channel. 

\subsection{Enhanced HARQ Model} 
Our proposal uses the MI to compute the number of bits to be transmitted at each transmission. After transmitting some bits of a given codeword, this codeword accumulates some mutual information. This mutual information can be computed knowing the number of bits transmitted and the attenuation coefficient affecting the transmitted bits. In this model we consider a reference $E_s/N_0$, which is a fixed value of $E_s/N_0$ in clear sky (for a given terminal and without attenuation). 
The model assumes that the channel is stationary for the transmission time of the bits, at a given transmission for a given codeword. The MI obtained at the $j^{th}$ transmission for a given codeword can be computed as:

\begin{equation}\label{MI_1}
\begin{aligned}
MI^{(j)}=N_{sent}^{(j)}.MI((\rho^{(j)})^2.\frac{E_s}{N_0}),
\end{aligned}
\end{equation}

where:
\begin{itemize}
\item $\rho^{(j)}$ is the attenuation coefficient affecting bits transmitted at the $j^{th}$ transmission for a given codeword;
\item $N_{sent}^{(j)}$ is the number of bits transmitted at the $j^{th}$ transmission;
\item $MI($\textperiodcentered$)$ is the function giving the value of mutual information for a given $E_s/N_0$ on a gaussian channel. 
\end{itemize}

The MI per bit accumulated for a given codeword, from the beginning of transmission until the $j^{th}$ transmission, can be computed as:
\begin{equation}\label{MI}
\begin{aligned}
MI^{(j)}_{acc}=\frac{N^{(j-1)}MI^{(j-1)}_{acc}+MI^{(j)}}{N^{(j)}},
\end{aligned}
\end{equation}

where:
\begin{itemize}
\item ${N^{(j)}}$ is the total number of bits transmitted for a codeword up to the $j^{th}$ transmission;
\item $MI^{(0)}_{acc}$=0.
\end{itemize}

Note that $\rho^{(j)}$ is unknown in our model, thus $MI^{(j)}_{acc}$ is also unknown.

Let us consider $MI_{needed}^{(j+1)}$ the minimum MI per bit needed to decode the codeword at the $(j+1)^{th}$ transmission with a predefined decoding probability. 

Let $N_{needed}^{(j+1)}$ be the number of bits to be transmitted at the $(j+1)^{th}$ transmission, we have:
\begin{equation}\label{needed_1}
\begin{aligned}
N_{bits}MI_{req}=N^{(j)}MI^{(j)}_{acc}+N_{needed}^{(j+1)}MI_{needed}^{(j+1)}\,,
\end{aligned}
\end{equation}

where $N_{bits}$ is the maximum number of bits that can be transmitted for a codeword in total. 

Finally $N_{needed}^{(j+1)}$ is given by:
\begin{equation}\label{needed_2}
\begin{aligned}
N_{needed}^{(j+1)}=\frac{N_{bits}MI_{req}-N^{(j)}MI^{(j)}_{acc}}{MI_{needed}^{(j+1)}}\,.
\end{aligned}
\end{equation}

$MI_{needed}^{(j+1)}$  and $MI^{(j)}_{acc}$ are the key parameters for the computation of the number of bits to be sent at the $(j+1)^{th}$ transmission. In the following, we explain in detail how to proceed to calculate these two values at each transmission according to the predefined decoding probabilities.

\subsection{Computation of $MI_{needed}^{(j+1)}$ and $MI^{(j)}_{acc}$}

In a first step, we will explain how to calculate $MI_{needed}^{(j+1)}$, then we finish by explaining the way we compute $MI^{(j)}_{acc}$.

We use the knowledge of the statistical distribution of the channel attenuation to control the probability of decoding a codeword at each transmission. 

The idea is to define at the beginning of the communication, a table containing the probability of decoding at each transmission. The decoding probability and the efficiency of the link usage are related, the sender can transmit a large number of bits at the first transmission, which increases the decoding probability but the efficiency will decrease and vice versa. So, we have to improve efficiency while respecting delay constraints for services. In addition the sender may also want to limit the number of transmission attempts for a given codeword. 

In the rest of this paper, we will consider $P_j$ the probability of decoding at the $j^{th}$ transmission conditioned on the fact that decoding at earlier transmissions was impossible, where $P=\sum_j P_j$ is the percentage of decoded codewords over all the transmitted codewords and is equal to 1. 


To target a decoding probability $P_j$ at the $j^{th}$ transmission, we have to find the corresponding $MI_{needed}^{(j)}$ necessary to calculate the number of bits $N_{needed}^{(j)}$ to be transmitted \eqref{needed_2}.  For this, we define $\rho_{needed}^{(j)}$ as the minimum successive attenuation coefficient (threshold) that provides a MI greater than $MI_{needed}^{(j)}$.  This threshold is the minimum attenuation needed to obtain the decoding probability at the $j^{th}$ transmission. $\rho_{needed}^{(j)}$ depends not only on the $j^{th}$ element in the probability decoding table but also on $\sum_{k=1}^{j-1}P_k$ as we will see later in \eqref{proba_decod}. $MI_{needed}^{(j)}$ is given by:

\begin{equation}\label{MI_req}
\begin{aligned}
MI_{needed}^{(j)}=MI((\rho^{(j)}_{needed})^2.\frac{E_s}{N_0}),
\end{aligned}
\end{equation}
Where $MI($\textperiodcentered$)$ is defined in \eqref{capacity_LMS}, that takes as input $(\rho^{(j)}_{needed})^2.\frac{E_s}{N_0}$. 

We assume that the reference $E_s/N_0$ fixed for a given terminal (only $\rho^2$ change over the time). $MI_{needed}^{(j)}$ depends only on the channel attenuation threshold $\rho_{needed}^{(j)}$. 
Since decoding probabilities at each transmission are predefined, $\rho_{needed}^{(j)}$ is pre-computed. $MI($\textperiodcentered$)$ is a strictly increasing function (as a function of $\rho$). Then any attenuation coefficient greater than $\rho_{needed}^{(j)}$ will lead to a successful decoding. 
To determine $\rho_{needed}^{(j)}$ leading to $P_j$, we use the cumulative distribution function (CDF) of the attenuations of LMS Channel. 

To simplify our calculation, we consider these two events:
\begin{itemize}
\item $A_j$: Successful decoding at the $j^{th}$ transmission;
\item $B_{j-1}$: Not decoding at the $(j-1)^{th}$ transmission.
\end{itemize}  
$P_j$ can be defined as $p(A_j\cap B_{j-1})$ and $p_j$ is $p(A_j)$. Since $A_j$ and $B_{j-1}$ are independent (according to the channel modelling),
\begin{equation}\label{proba_decod}
\begin{aligned}
P_j=p_j(1-\sum_{k=1}^{j-1}P_k),\\
p_j=\frac{P_j}{(1-\sum_{k=1}^{j-1}P_k)}\,.
\end{aligned}
\end{equation}

\begin{figure}[b]
\centering
\captionsetup{justification=centering}
\includegraphics[scale=0.37]{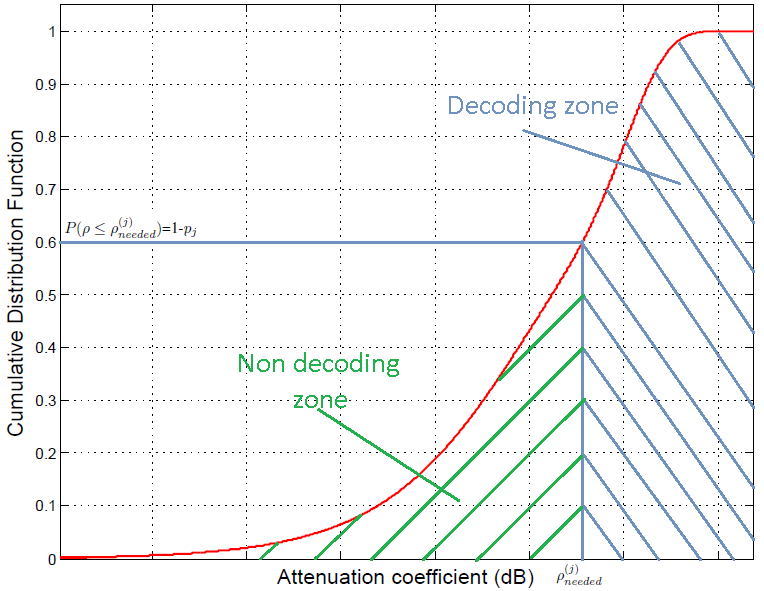}
\caption{CDF of the attenuation coefficients in the LMS channel.}
\label{fig:Cdf_its} 
\vspace{-15pt}
\end{figure}

CDF of the channel gives us $P(\rho \leq \rho_{needed}^{(j)})$, while $p_j$ corresponds to $P(\rho \geq \rho_{needed}^{(j)})$ (successful decoding). Therefore, $\rho^{(j)}_{needed}$ on the CDF graph is given by 1-$p_j$ (See Figure \ref{fig:Cdf_its}). 

Once we found $\rho^{(j)}_{needed}$ leading to $P_j$, we use it in \eqref{MI_req} to calculate $MI_{needed}^{(j)}$. 

At the $(j-1)^{th}$ transmission, only codewords affected by attenuation coefficients greater than $\rho^{(j-1)}_{needed}$ are decoded. $MI^{(j-1)}_{acc}$ represents the average mutual information per bit of all codewords that have not been decoded at the $(j-1)^{th}$ transmission, affected by attenuation coefficients less than $\rho^{(j-1)}_{needed}$. $MI^{(j-1)}_{acc}$ is the sum of all mutual informations weighted by their probabilities in the non decoding zone (see Figure \ref{fig:Cdf_its}). For this, we use the CDF and the PDF (probability density function) of the channel. $MI^{(j-1)}_{acc}$ can be calculated by:
\begin{equation}\label{mean_mi}
\begin{aligned}
MI^{(j-1)}_{acc} = \sum_i p_{wi} MI((\rho_i)^2.\frac{E_s}{N_0}),
\end{aligned}
\end{equation}
where:
\begin{itemize}
\item $\rho_i$ belongs to the ensemble of attenuation coefficients, which are less than $\rho^{(j-1)}_{needed}$;
\item $p_{wi}$ is the weighted probability, defined as the probability to obtain $\rho_i$ in the non decoding zone.
\end{itemize}

Finally, to calculate $N_{needed}^{(j)}$ we use $MI_{needed}^{(j)}$ and $MI^{(j-1)}_{acc}$ in \eqref{needed_2}. Where $N^{(j-1)}$ is the sum of all numbers of bits already calculated for the previous transmissions.

\begin{figure}[t]
\centering
\captionsetup{justification=centering}
\includegraphics[scale=0.4]{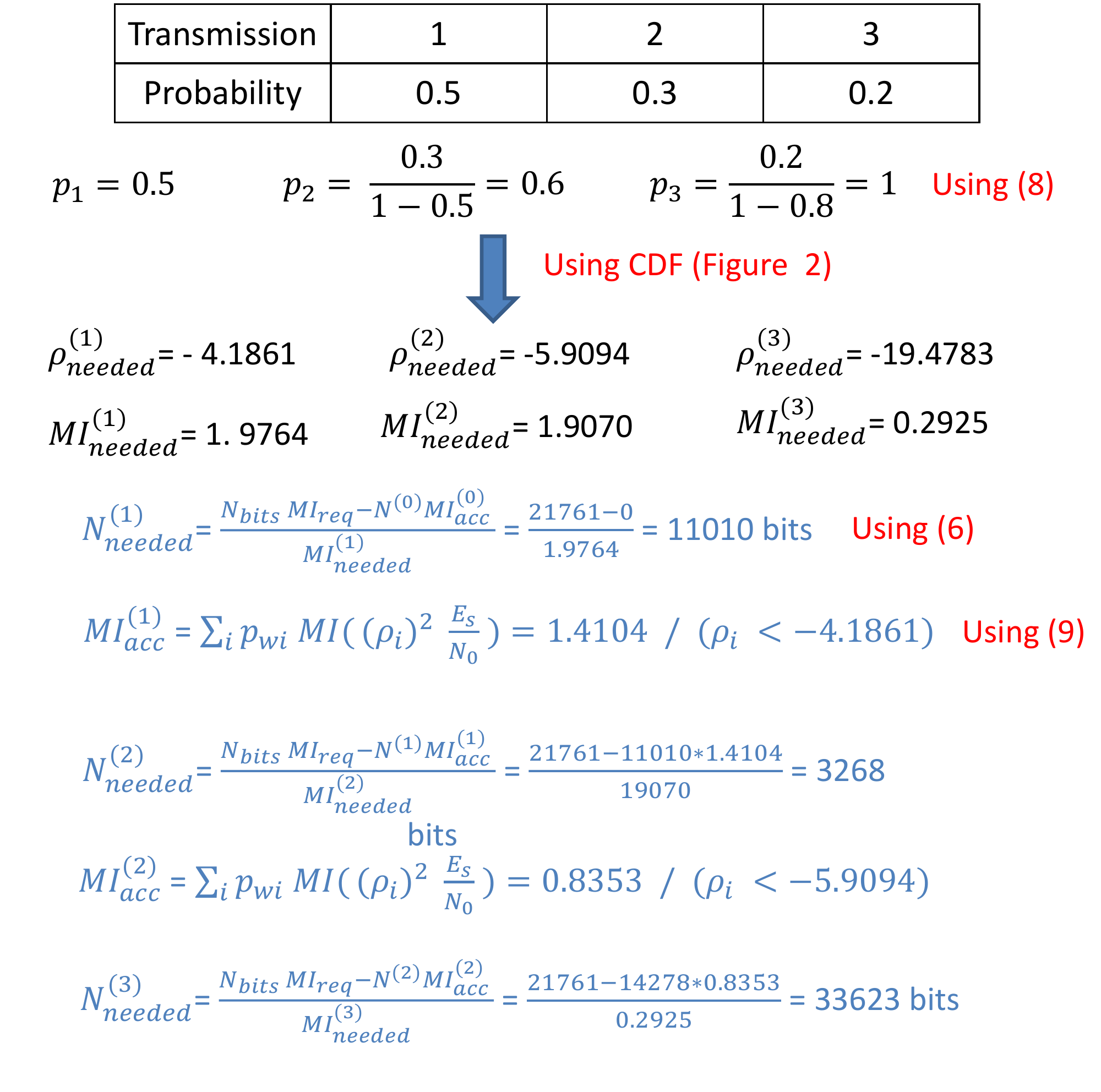}
\caption{Numerical example for the computation of number of bits to be transmitted at each transmission ($E_s/N_0 = 13 \,dB$, code (8920,1/6), three transmissions).}
\label{fig:example_trans} 
\vspace{-15pt}
\end{figure}

Figure \ref{fig:example_trans} represents a numerical example for computation of the numbers of bits to be transmitted at each transmission, showing different steps for the calculation of all parameters in our model.



\section{Simulations and Results} \label{sec:res}
The system parameters that we will consider for all simulations are:
\begin{itemize}
\item Satellite Orbit: Geostationary (GEO)
\item Round Trip Time: 500 ms
\item Band: S
\item Land mobile satellite channel , Intermediate Tree Shadowed Environment (ITS)
\item Speed: 60 Km/h
\item Distance: 10 Km
\item Mother FEC code, CCSDS Turbo Codes 1/6
\item Codeword length: 53520 bits (Data bits : 8920 bits)
\item Modulation: QPSK
\item Symbol time : $4.10^{-6}$ seconds, bit rate (Rb): 500 Kbps 
\end{itemize}
Our simulations are about 10 minutes of communication between the transmitter and the receiver (about 300 Mb transmitted). In our simulations, we consider a targeted WER of $10^{-4}$, and we use the actual performances of CCSDS Turbo codes (8920,$\frac{1}{6}$) as presented in \cite{Dolinar}. Attenuations in ITS environment are very high, which makes difficult the decoding of 100\% of codewords in the range of reference $E_s/N_0$ considered (7 to 13 dB) with code rates greater than $\frac{1}{6}$. Hence the choice of code rate $\frac{1}{6}$. We suppose that the synchronisation is never lost. 
The transmitted codewords are identified with a sequence number that is never lost. We suppose also that the return channel does not introduce errors and the feedback can be transmitted immediately (no congestion problem on the reverse link).

A calibration phase is required in the simulations, that take into account the actual numerical performances of the targeted FEC code(s). This allows to avoid implementing a real decoder in the simulation chain, while assuring a very good accuracy of the representation \cite{predic_perf}. We use the MI to decide if the codeword is decoded or not using this formula:
\begin{equation}\label{decod}
\begin{aligned}
N^{(j)}MI^{(j)}_{acc}\ge N_{bits}MI_{req}\,.
\end{aligned}
\end{equation}

\vspace{0pt} 

\subsection*{Simulations of classical IR and proposed enhanced HARQ}
We present results obtained by implementing both schemes (classical IR and proposed enhanced HARQ) described in previous sections, and we compare these results. 

We define a set of precomputed decoding probabilities, that provide for each  retransmission a precomputed number of parity bits to send. We are interested to improve the efficiency while maintaining an acceptable delay for services. 


We define the efficiency, E (bits/symbol), as follows:
  \begin{equation}\label{spectral_efficiency}
\begin{aligned}
E=\frac{N_{data}\,\,_{bits} N_{decoded}\,\,_{words}}{N_{total}},
\end{aligned}
\end{equation}
where: 
\begin{itemize}
\item $N_{data}\,\,_{bits}$ is the number of data bits (useful bits) per codeword considered in our coding scheme;
\item $N_{decoded}\,\,_{words}$ is the total number of decoded codewords during the communication;
\item $N_{total}$ is the total number of symbols transmitted during the communication.
\end{itemize}

The delay for decoded codewords (at the receiver) can be expressed in terms of number of transmissions ($N_{trans}$), bit rate ($R_b$), number of bits sent ($N$) and propagation delay ($T_{propag}$), assuming a negligible access delay: 
\begin{equation}\label{delay}
\begin{aligned}
Delay= \frac{N}{R_b}+2(N_{trans}-1)T_{propag}+T_{propag}\,\,\, (s)\,.
\end{aligned}
\end{equation}

The efficiency and the delay are positively related. To improve the efficiency we can increase the delay while respecting the delay constraints for services.
This delay is controlled by the predefined decoding probabilities, which will be quasi constant along all the values of reference $E_s/N_0$. 
In the following, we have considered three different tables of decoding probabilities. 
Fixed decoding probabilities at each transmission considered in our simulations are given in Table \ref{table:proba_threecases}, where the maximum number of transmissions for a given codeword is four. Case 1 corresponds to a service accepting the delivery of 80\% of the messages at the first two transmissions and 20\% at the last two retransmissions. Case 2 corresponds to a service accepting the delivery of 95\% of the messages at the first two transmissions and 5\% at the last two retransmissions. Case 3 corresponds to a service accepting the delivery of 45\% of the messages at the first two transmissions and 55\% at the last two retransmissions. 
\begin {table}[h!]
\begin{center}
\captionsetup{justification=centering}
\caption{PREDEFINED DECODING PROBABILITY TABLE FOR THE 3 CONSIDERED CASES}
\begin{tabular}{|c|c|c|c|c|}
  \hline
  Transmission & $1^{st}$ & $2^{nd}$ & $3^{rd}$  & $4^{th}$  \\
  \hline 
 $P_i (case 1)$ & $0.5$ & $0.3$ & $0.15$ & $0.04999$  \\
  \hline
  $P_i (case 2)$ & $0.75$ & $0.2$ & $0.03$ & $0.01999$  \\
  \hline
  $P_i (case 3)$ & $0.4$ & $0.05$ & $0.45$ & $0.0999$  \\
  \hline
\end{tabular}
\label{table:proba_threecases}
\vspace{-10pt}
\end{center}
\end {table}


The number of bits to be transmitted at each transmission for the classical IR HARQ scheme (Section III) can follow several strategies. As a simple case, we consider an equally shared repartition of the data+parity bits among the transmissions, as shown in Table \ref{table:static}.

\begin {table}[h!]
\vspace{-1pt}
\begin{center}
\captionsetup{justification=centering}
\caption{NUMBER OF BITS TO BE TRANSMITTED AT EACH TRANSMISSION FOR CLASSICAL IR HARQ MODEL}
\begin{tabular}{|c|c|c|c|c|}
  \hline
  Transmission & $1^{st}$ & $2^{nd}$ & $3^{rd}$  & $4^{th}$  \\
  \hline 
 $N^j_{sent} (bits)$ & $13380$ & $13380$ & $13380$ & $13380$  \\
  \hline
\end{tabular}
\label{table:static}
\vspace{-10pt}
\end{center}
\end {table}

After simulating first (classical IR HARQ) and second (enhanced approach with three cases mentioned above) schemes, under the same conditions, we figure out the efficiency and the delay obtained for both schemes and compare them.

\begin{figure}[t]
\centering
\captionsetup{justification=centering}
\includegraphics[scale=0.56]{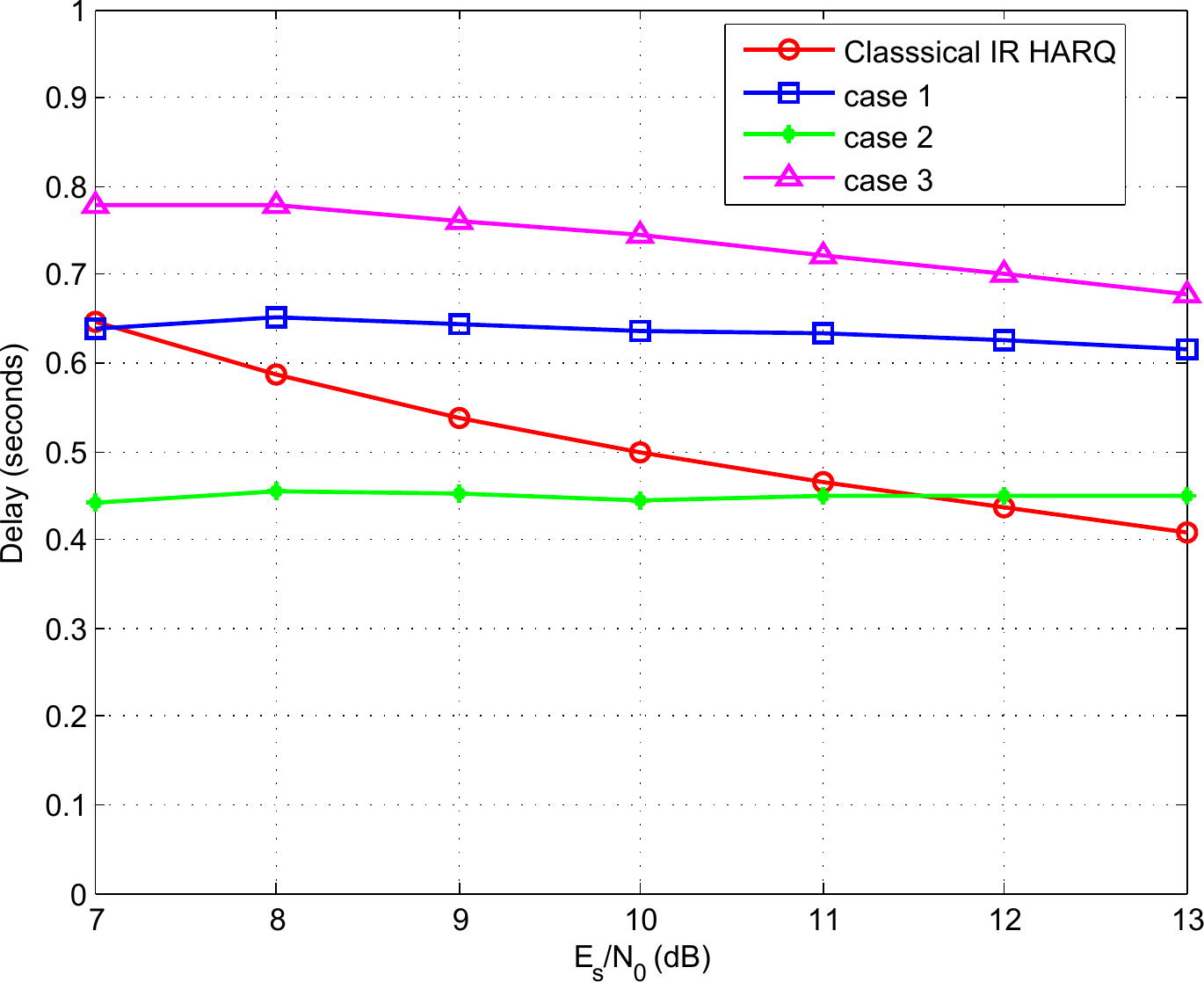} 
\caption{Delay obtained for both models for different values of reference $(E_s/N_0)_{dB}$.}
\label{figure delai}
\vspace{-15pt}
\end{figure}

Figure \ref{figure delai} shows the average delay, required to decode codewords, obtained with both models (by means of simulations). The mean delay is computed by averaging delays obtained for decoded codewords calculated using \eqref{delay}. As we can see, the values obtained for each case of the proposed model are approximately the same. The delay remains stable and controlled by the decoding probability, globally constant. It changes a little, from a reference $E_s/N_0$ to another, according to the number of bits transmitted and the values of decoding probabilities.

\begin{figure}[t]
\centering
\captionsetup{justification=centering}
\includegraphics[scale=0.56]{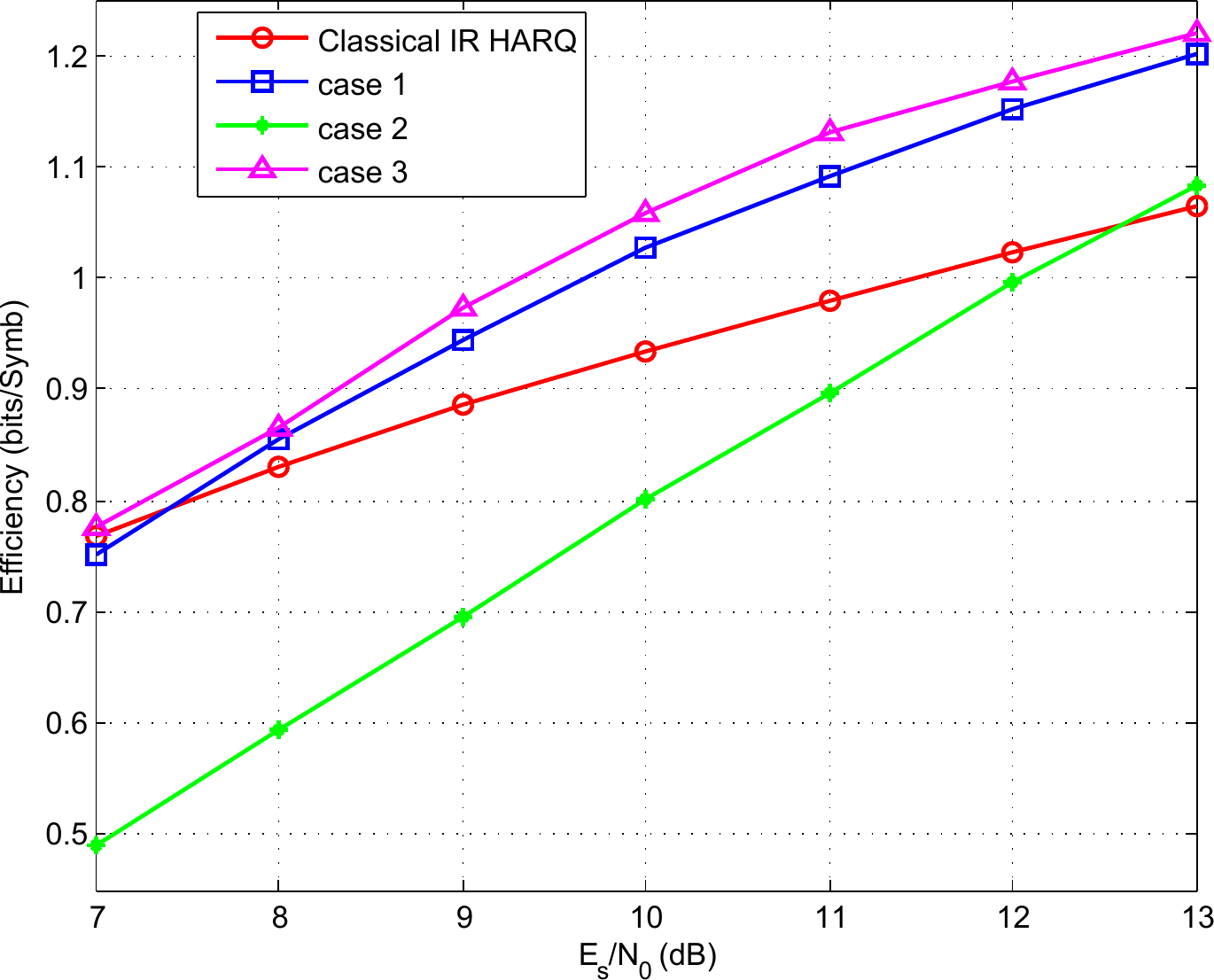} 
\caption{Efficiency obtained for both models for different values of reference $(E_s/N_0)_{dB}$.}
\label{fig:eff}
\vspace{-15pt}
\end{figure}

Figure \ref{fig:eff} compares the efficiency obtained with both models in the same conditions. We can see that even if we loose in delay, the proposed model in cases 1 and 2, outperforms the classical one especially for high values of reference $E_s/N_0$. The resulting gain can reach 15\%. These results validate our proposal, that fixing decoding probabilities at each transmission can improve the efficiency while respecting delay constraints for a given service. 
These results seem promising, however they have been obtained without any optimization of decoding probability for each transmission step. Therefore, some further improvements could probably be obtained. This clearly calls for an optimization process in further steps of the work.

\section{Conclusion and Future Work} \label{sec:con}
In this paper, we  have compared two techniques of HARQ transmission. The first one is a classical IR HARQ scheme, that transmits the same number of bits at each transmission without any knowledge about the global statistics of the channel; the second one is an enhanced HARQ technique, which takes into account the global statistics of the channel. This proposed enhanced technique estimates the number of bits to be transmitted at each transmission to decode a codeword with a given probability. This estimation relies on the mutual information given the decoding probability at each transmission and the knowledge of the distribution of the channel attenuations. Finally, results obtained after simulating both schemes in a mobile satellite communication environment are compared in terms of decoding probability (delay) and efficiency. Results show that enhanced HARQ has better performance in terms of efficiency especially for high values of reference SNR, while maintaining a quasi constant delay acceptable by delay constraints of service. 
As future work we will modify the proposed method to make it an adaptive model. This model would calculate the number of bits to be transmitted  at each transmission according to decoding probabilities and real values of attenuation coefficients obtained. We plan also to consider additional system parameters, such as framing constraints and overhead to evaluate the performance of the mechanism in real systems.

\bibliographystyle{ieeetran}
\bibliography{mybib}

\end{document}